# Welding few-layered graphene ribbon via penetration of high-speed fullerenes

Jiao Shi [1*], Chunwei Hu [1], Jiaxing Ji [1], Bo Song [1*]

[1] College of Water Resources and Architectural Engineering, Northwest A&F University, Yangling 712100, China
*Corresponding author
Email addresses: shijiaoau@163.com (J. Shi); songbo943@163.com (B. Song)

**Abstract:** Two-dimensional materials are popular in design of nanodevices. To improve bending stiffness of few-layered graphene ribbon, the relative sliding between neighboring layers should be avoided by method such as bonding them together. In this study, the new bonds between the graphene layers in ribbon are formed by ballistic $C_{60}$ fullerenes from both surfaces of the ribbon. Collision effects are evaluated through molecular dynamics simulations. The formation of the new two dimensional carbon material is demonstrated by breakage and generation of bonds between atoms from graphene and/or fullerenes. Note that the incident velocity of the fullerenes had an interval, in which both graphene and fullerenes will break and further be connected by new covalent bonds at the impact area.

## 1. Introduction

Among all the carbon allotropes, graphene is the most popular low-dimensional material over the last 16 years [1]. Benefitting from the peculiar electron configuration of carbon atom, graphene exhibits excellent mechanical properties [2, 3], electrical properties [4, 5] and thermal conductivities [6], and has been applied in varies types of engineering [7-9]. Besides, owing to massive production with low cost, multi-layered graphene ribbon is more popular in engineering than the single layer sheet. Although the in-plane mechanical properties, e.g., modulus and strength, of graphene ribbon is excellent, its out-of-plane bending stiffness is very low [10, 11]. For multi-layered graphene ribbon, the super-lubrication between the adjacent layers leads to easily relative sliding, which reduces its out-of-bending stiffness, and therefore, confines its potential application in the nano-electro-mechanical system (NEMS).

To improve its bending stiffness, structural modifications on the prestine multi-layered graphene ribbon are required. For example, people developed diamond film [12-14]. Such $sp^2/sp^3$ composited material has excellent mechanical properties both in-plane and out-of-plane. By putting two or more layers of graphene at high pressure, Martins *et al*. [15] tested the phase transition that leading to diamondene. The thermal and mechanical properties of the nanotube that made from diamondene were also predicted [16-18]. Chernozatonskii *et al*. [19] studied diamane ($C_2H$ layer) of its mechanical and electrical properties [20, 21]. Jiang *et al*. [22] theoretically proposed a new double-layered carbon allotrope that has strain-dependent bandgap. Yang *et al*. [23, 24] formed nanotextures from graphene ribbons. However, large-scale synthesis of these structures is hindered by the high pressures needed to initiate the graphene-diamond phase transformation. Apart from the above graphite-based materials, new 2D carbon materials based on other low dimensional carbon materials, such as carbon nanotube [25-27] and fullerenes [28], were also proposed. For instance, Baughman and Galvão [29] presented a carbon network that has auxetic property. Hall *et al*. [30] suggested a model of buckypaper from carbon nanotubes. Xu *et al*. [31] used graphene and carbon nanotubes to form a three dimensional network.

Meanwhile, there are other techniques, e.g., irradiation methods [32-36] and shot peening methods [37, 38], were also introduced to form new carbon structures from graphene. By using oblique ion beam irradiation, Bai et al [33] obtained many types of nano-porous graphene with tunable geometries. Using molecular dynamics (MD) method, Abdol et al [32] built a three-dimensional graphene networks via silicon ion beam irradiation. In a series reports by Sadeghzadeh [34, 39-42], the graphene sheets were impacted with various kinds of high speed projectiles, and the effects of several parameters on the rupture form and penetration-resistance efficiency of the graphene ribbons were evaluated systematically. In the experiments by Eller *et al.* [43], few layered graphene were bombarded with golden nanoparticles, and the ejection of ions from graphene were characterized. Similarly, the ions and electrons emitted from a keV fullerene after colliding a few-layered graphene were studied both experimentally and numerically [44, 45]. In simulations, adaptive intermolecular reactive bond order (AIREBO) potential [46] was used to describe the interaction among carbon atoms. Golunski and Postawa [47] discussed the effect of initial speed and impact angle of keV $C_{60}$ on emission of carbon atoms from graphene after collision. Using MD method and AIREBO potential, Becton *et al*. [48] proposed a programmable design scheme of nano-porous graphene by impacting ideal graphene with high speed fullerenes, in which the sizes and layout of the pores could be well controlled. Liesegang and Oligschleger [49] investigated the density of states of the graphene structures grafted with fullerenes. Hosseini-Hashemi *et al*. [50] investigated the vibration of a graphene impacted by a fullerene using both analytical model and MD simulations. Combined ab initio calculation with finite element simulation, Signetti *et al*. [51] study the ballistic properties of 2D materials (boron nitride, graphene, or alternated graphene and boron nitride) under impact of extremely high speed $C_{60}$. From this research, design of 2D materials-based nanocomposites for armors was enhanced.

Despite the recent success in impacting graphene by irradiation method or shot peening method, most of them aim to protect the structures and devices from the penetration of high-energy impacting projectiles. Obviously, little attention has been paid on the carbon network that formed after the collision, not to mention the quantitatively assessment of the bonding changes at an impact area. When a multi-layered graphene ribbon is penetrated by fullerenes, many carbon-carbon bonds in the graphene will break, while some new bonds may generate between the collision-induced unsaturated atoms. The newly formed bonds at the impacted area may form a new carbon network, which will connect the neighboring layers in the ribbon.

In this work, we adopt the shot peening method to form a new 2D carbon material from multi-layered graphene ribbons, which intends to prevent the relative sliding between neighboring layers and improve the out-of-plane bending stiffness of graphene ribbon. $C_{60}$ fullerenes are introduced as shot blast stream to "weld" the layers of graphene in ribbon from both sides. The injection velocity of fullerene that can penetrate the ribbon is estimated with respect to one to four layers of graphene ribbons at 300 K. In particularly, the bond breakage and generation during collision are will be assessed. A fundamental understanding of the formation mechanism for the carbon network will be analyzed.

## 2. Model and methods

2.1 model

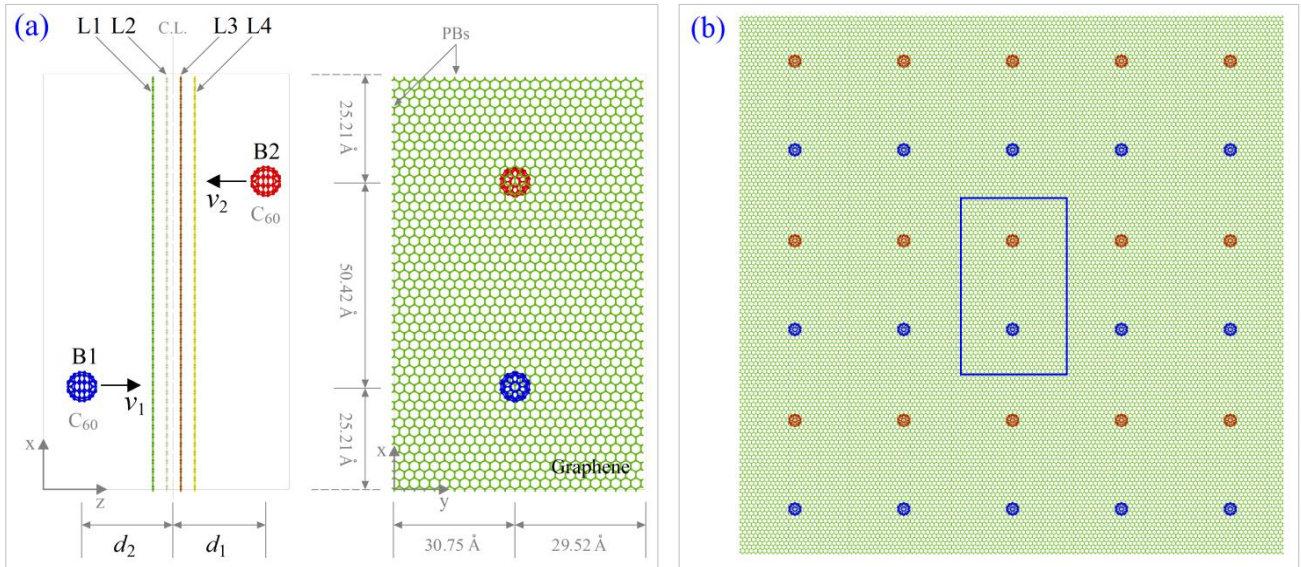

Figure 1 Schematic of shot peening to a few-layered graphene ribbon. (a) In a periodic cell, an $m$-layered graphene ribbon with A-A stack is subjected to collision by two fullerenes (namely B1 and B2) from both sides. High speed fullerenes that play as shot blast stream have the same incident velocity (i.e., $v_1=v_2$, perpendicular to the ribbon) and the same initial distance with the central layer (C.L.) of the ribbon (i.e., $d_1=d_2 > 20$ Å). (b) A ribbon with 5×3 periodic cells.

In Figure 1a, a model with $m=4$ for shot peening test is given. Each layer of the ribbon contains 2400 carbon atoms. In simulation, $m=1, 2, 3$ and $4$ are involved. When $m=1$, only Layer 1 (L1) exists in the model. Similarly, when $m=2$, L1 and L2 exist. When $m=3$, L1, L2 and L3 exist. The initial distance between the adjacent graphene layers is identical to 0.34 nm.

The injection velocities of fullerenes are specified symmetrically and simultaneously. In this study, one purpose is to find **the injection velocity interval of the nanoballs when impacting a graphene ribbon**. For example, the lower boundary of the interval is set as the minimal incident velocity at which the two fullerenes covalently bond with the graphene after collision. **With injection velocity smaller than that, the fullerenes will be bounced back from or just attracted upon the graphene ribbon** via the non-bonding interaction. However, **if the incident velocity is higher than the upper boundary of the interval, the nanoballs may penetrate the ribbon and some carbon atoms escape from the structure**.

To form a network during collision, some bonds in the nanoballs and the graphene ribbons need to be broken, and then some of the unsaturated atoms will bond together, i.e., forming new bonds. When the neighboring layers in the multi-layered ribbons are connected by the new carbon-carbon bonds, a new 2D carbon network is obtained. Hence, another task of present study is to **evaluate the collision-induced damage and quantitatively reveal the bond breakage and generation between nanoballs and graphene ribbons during the collision**.

2.2 Methodology

The injection velocity interval with respective to a graphene ribbon with given layers is searched from the initial interval of [0, 200] Å/ps (1 Å/ps =0.1 km/s) by bi-sectioning method. Besides, the accuracy of the interval is set as 5 Å/ps with consideration of the influence of randomly thermal vibration of atoms in the system.

To reflect the state of the graphene ribbon during the collision, the variation of potential energy (VPE) of ribbon is recorded by subtracting the current potential energy from its initial value, i.e.,

$$\text{VPE}(t) = \text{PE}(t) - \text{PE}(t_0), \quad (1)$$

where PE($t$) and PE($t_0$) are the potential energy of the graphene ribbon at times $t$ and $t_0$, respectively. The value of PE($t$) can be obtained by substituting the positions of atoms in the graphene ribbon at time $t$ into the AIREBO potential function.

During the collision, the configuration of fullerenes and the graphene ribbons as well as the detailed information of bond breakage and formation between them is visualized and counted by the Ovito [52], a type of visualization software for post-processing.

To obtain the velocity interval and the responses of the ribbon and the nanoballs, molecular dynamics simulation approach is adopted. The simulations are carried out on the large-scale atomic/molecular massively parallel simulator (LAMMPS) [53]. Considering electric neutrality, the interaction between the carbon atoms is evaluated by AIREBO potential [46], which contains three items, i.e., item one is the reactive empirical bond order (REBO) part for describing the bond-ordered short-range interaction, item two is the torsion part to consider the dihedral effect in the local deformation, and the final item is the Lennard–Jones (L-J) potential [54] to describe the long-range non-bonding interaction between two atoms with a distance smaller than 1.02 nm.

In each simulation, there are four major steps:

(1) Build the initial model with specified components (e.g., an $m$-layered ribbon and two fullerenes) and given geometry as shown in Figure 1a.

(2) Fix the two fullerenes, and relax the graphene ribbon in the canonical (NVT: constant Number of atoms, constant Volume and Temperature of system) ensemble for 200 ps with timestep of 0.001 ps. The temperature is controlled to be 300 K by the Nose-Hoover thermostat [55, 56].

(3) After relaxation, two fullerenes start to move toward the ribbon with the same speed. The collision process with duration of 3 ps is simulated at NVE (constant Number of atoms, constant Volume and Energy of system) ensemble. To precisely reveal the collision process, the timestep is reduced to 0.05 fs.

(4) After collision, the whole system will be relaxed in an NVT ensemble again for 300 ps. The timestep is 1.0 fs, and the temperature is still 300 K.

## 3. Numerical tests and discussion

### 3.1. The critical intervals of injection velocities of fullerenes

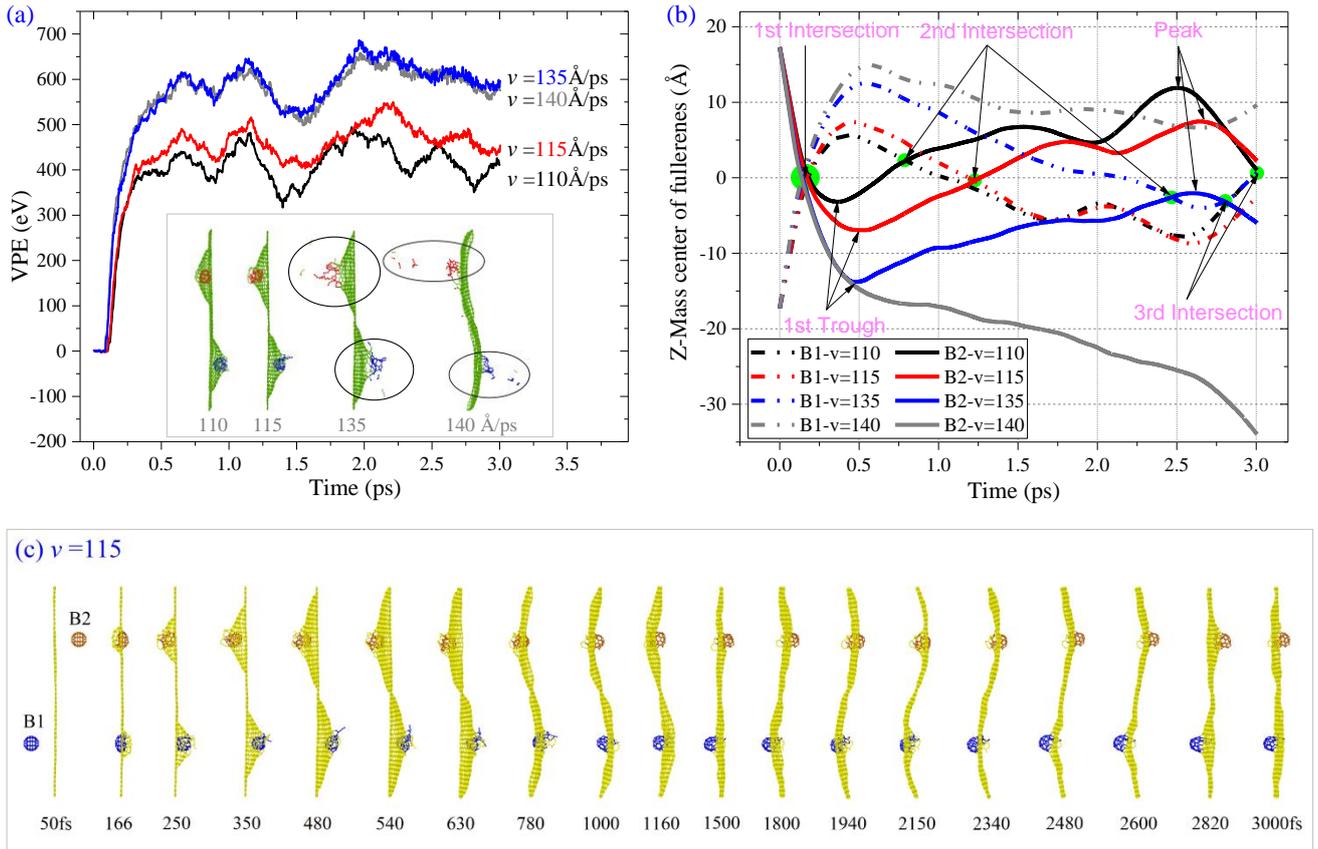

Figure 2  Physical properties of the single-layered graphene ribbon ($m$=1) collided by two fullerenes at 300 K. (a) Variation of potential energy (VPE) of graphene L1. (b) Traces of the two fullerenes in z-direction. (c) Snapshots of the system during collision with $v$=115 Å/ps.

Using bi-sectioning method, the intervals of injection velocities for the four models are obtained. When the injection velocities of the two fullerenes are in the interval, the fullerenes will "fuse" onto the ribbons and fail to escape. Briefly, when $m$=1, i.e., single layer of graphene impacted, the interval converges at [115, 135] Å/ps. For $m$=2, the interval is [130, 165] Å/ps. The interval is [140, 200] Å/ps for $m$=3, and [160, 260] Å/ps for $m$=4. Here, both the lower and upper limits of interval increase when the ribbon contains more layers of graphene. The reason is that with more layers support each other during collision, the fullerenes need higher initial kinetic energy to break them.

Results also indicate that the lower and upper limits of an interval are determined by the both states of fullerenes and graphene after collision. When $m$=1, the dynamic responses of the two fullerenes and graphene L1 are illustrated in Figure 2. In Figure 2a, large amplitude of fluctuation of the VPE curves is caused by two reasons. One is the bond breakage and generation during collision. The other is mechanical vibration of the damaged graphene during collision. Besides, the peak value of VPE with respect to $v$=115 Å/ps is higher than that of $v$=110 Å/ps but lower than those of $v$=135 and 140 Å/ps. From the inserted snapshots in Figure 2a, we find that the two fullerenes are just bonded slightly with the damaged graphene after collision with $v$=110 Å/ps. However, the nanoballs break after collision and the graphene is damaged heavier when $v$=115 Å/ps (Movie 1). Hence, higher incident velocity of fullerenes leads to heavier damage of graphene and nanoballs after collision, and further causes higher peak value of VPE. The VPE curves with respect to $v$=135 and 140 Å/ps differ slightly, which indicates that the graphene ribbons have the same level of damage after collisions.

In Figure 2b, the z-distances of the two fullerenes are drawn at four different cases. For each collision, the two curves have the same color. As the two balls have the same initial distances ($d_1=d_2$) away from the graphene, the first intersection happens at z=0 for each case. But later, the four pairs of curves show different characteristics. For example, the first troughs in the four solid lines, which means the fullerenes have the maximum displacement along their injection paths (480 fs in Figure 2c), appear later at higher incident velocity.

After approaching the maximum displacement, the balls start to bounce back from the graphene. Except the case with respect to $v$=140 Å/ps, each of the rest curves has a second intersection, which means that the graphene bonded with fullerene bounces back for the first time to the initially equilibrium position (z=0) (1160 fs in Figure 2c). It is found that the second intersection happens later at higher injection velocity. Like moving further needs longer time, bouncing back also requires more time. The damaged graphene together with fullerenes does not stop after approaching at z=0, because the atoms on fullerenes have collected translational kinetic energy during the bouncing back process. Therefore, the atoms keep moving till they have the maximal opposite displacements (2480 fs in Figure 2c). After that, they bounce back again. But this time, their velocities have the same direction as the related fullerene's initial injection velocity (2600 fs in Figure 2c). As they meet at z=0 again, a period of mechanical vibration of graphene is completed. Clearly, fluctuation of the trace curves in Figure 2b indicates that the fullerenes are not bouncing synchronously with their connected part of the graphene. This can be verified by the difference of the snapshots with respect to 1940 fs, 2150 fs, and 2340 fs in Figure 2c.

Figure 2b also indicates that the period of vibration of the graphene with respect to $v$=110 Å/ps is ~3 ps, which is lower than that with respect to $v$=115 Å/ps. But the period becomes shorter when $v$=135 Å/ps. The reason is that the graphene and the fullerenes are damaged seriously. For the case of $v$=140 Å/ps, the solid curve decrease continuously because some atoms on B2 escape from the graphene after collision.

3.2. Bond breakage and generation during collision

*(a) Bonds between atoms in graphene and fullerenes*

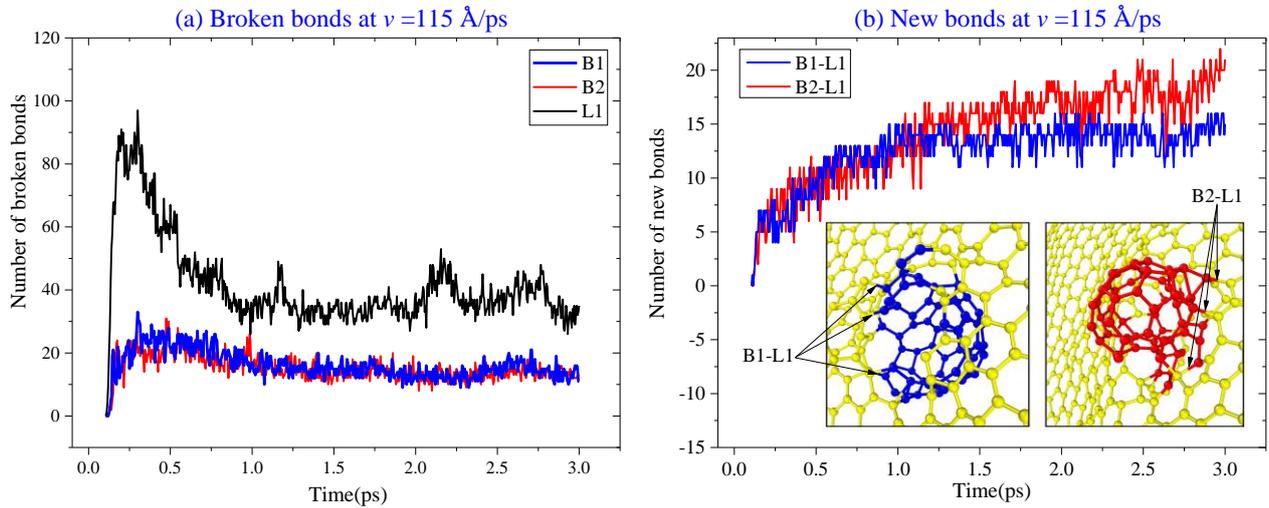

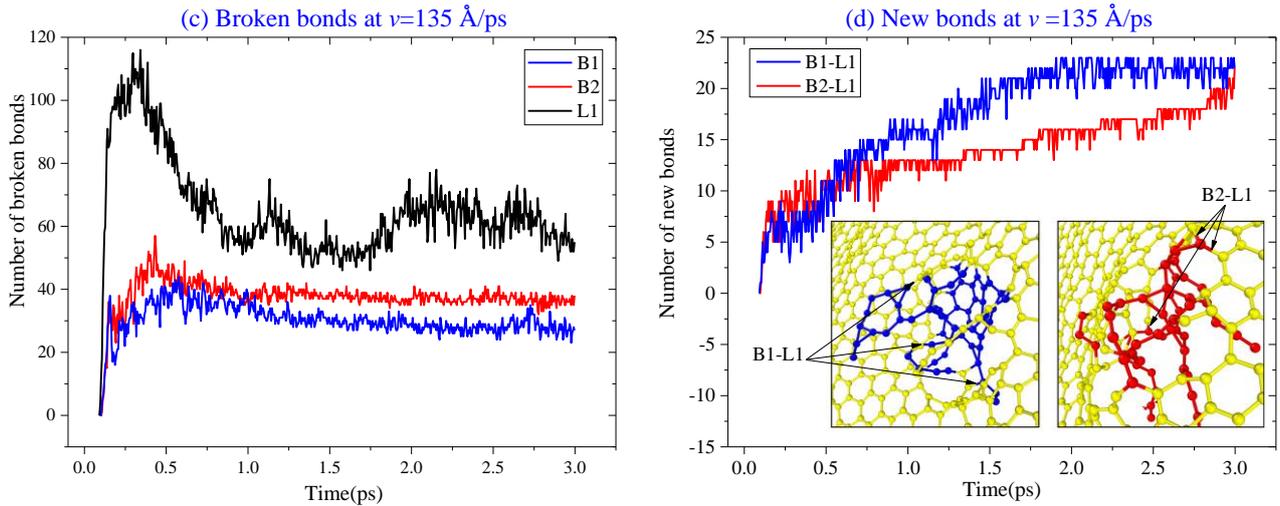

Figure 3  Histories of bond number in single-layered graphene ribbon system. (a) The numbers of broken bonds in graphene (L1) and fullerenes (B1 and B2) with $v$ =115 Å/ps. (b) Generation of new bonds between fullerenes and graphene when $v$ =115 Å/ps. (c) The numbers of broken bonds in graphene and fullerenes when $v$ =135 Å/ps. (d) Generation of new bonds between fullerenes and graphene when $v$ =135 Å/ps. Snapshots of the two impact areas at 3 ps are inserted in (b) and (d). The cutoff for bond length is 1.65 Å, i.e., two atoms with distance higher than the cutoff cannot covalently bonded.

As aforementioned, the lower and upper limits of the incident velocity of the two fullerenes are 115 Å/ps and 135 Å/ps, respectively, in collision with a single-layered graphene ribbon. To show the collision effect, statistics of the broken bonds in the three components and the new bonds are given in Figure 3. Considering stable and strong interaction between two atoms, the cutoff of bonds in statistics is set to be 1.65 Å. This criterion will be adopted throughout the paper, unless otherwise stated.

When the nanoballs have the lower speed, i.e., 115 Å/ps, to collide with the graphene from both sides simultaneously, both fullerenes and graphene are broken, and they are finally connected by covalent bonds. The number of broken bonds in graphene L1 is between 35 and 40 during collision (Figure 3a). During the period, each fullerenes have almost the same number of broken bonds, approximately 16. Broken bonds in the three components provide opportunity for generation of new bonds between neighboring unsaturated atoms. Figure 3b indicates that the two broken balls have different number of covalent bonds with graphene. The number of new bonds between B1 and L1 is about 15, which is lower than that (~20) between B2 and L1. The summation of new bonds is approximately equal to the broken bonds in L1. **From the inserted snapshots in Figure 3b, the local bond configurations at the two areas are also different. The reason is that the thermal vibration of atoms on graphene lead to asymmetric collision with the two nanoballs. Hence, one can conclude that synchronous collision by two nanoballs does not mean that they will have identical connections with the graphene**.

If improving the incident velocity to the upper limit, i.e., 135 Å/ps, the three components have higher number of broken bonds than at lower speed collision. It is reasonable that stronger collision lead to heavier damage of the three components. For example, the broken bonds in graphene L1 is higher than 50. Fullerene B1 has ~27 broken bonds, but B2 has ~38 broken bonds after collision (Figure 3c). Most atoms in the two seriously damaged fullerenes are captured by the graphene L1 with larger number of covalent bonds (Figure 3d). The total number of new bonds is lower than the number of broken bonds in L1. It says that the graphene is seriously damaged, which can be verified by the snapshots at two impact areas at 3 ps. Dangling bonds can be found in the system after collision. Especially, from the snapshot with respect to "135" in Figure 2a, some atoms on graphene escape from the system after collision.

*(b) Bonds between atoms in graphene*

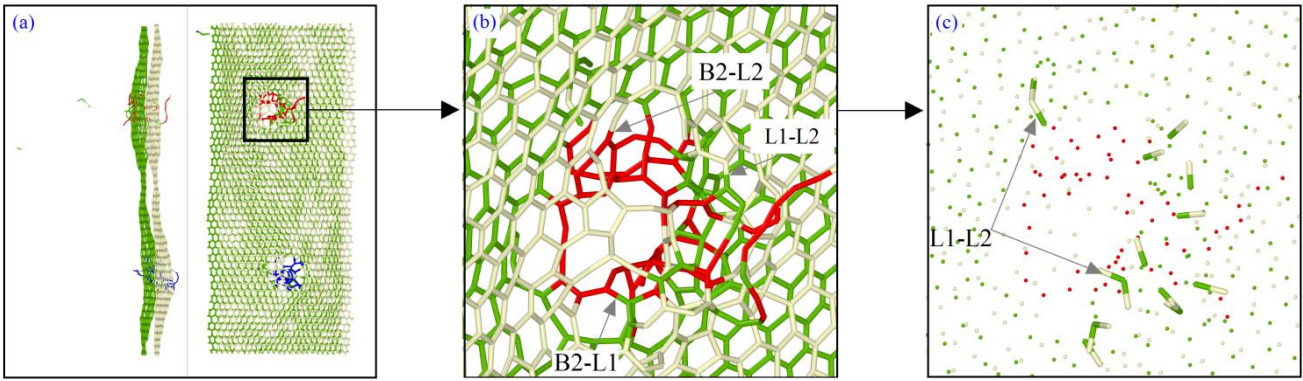

Figure 4  Carbon network, i.e., bond layout (at 3 ps), of the system with two layers of graphene (L1, L2) collided by fullerenes (B1, B2) with $v$ =165 Å/ps. (a) Side- and oblique-views of the ribbon. (b) The B2 impact area. (c) New bonds at the B2 impact area. After collision, some fullerene atoms bond with the atoms from the damaged graphene. B2-L1 and B2-L2 mean the bonds between atoms on B2 and L1 or on B2 and L2, respectively. L1-L2 means bond between the two layers of graphene.

When the multi-layered graphene ribbon is impacted by high-speed fullerenes, new bonds are generated between atoms in neighboring layers. For example, using two fullerenes with the same incident velocity of 165 Å/ps to collide with the two-layered graphene ribbon, the two layers are bonded after collision. The snapshot at 3 ps as shown in Figure 4 is chosen to show the layout of bonds at impacted areas. The global configuration in Figure 4a indicates that the two nanoballs penetrate the two layers of graphene, and some atoms escape from them after collision. By magnifying the local area collided by B2, we find that both layers of graphene and B2 have been seriously damaged and they are connected by covalent bonds of B2-L1, B2-L2 and L1-L2. Most atoms on fullerene B2 are between the two layers of graphene. Hence, the two layers can be considered as being "welded" by fullerene. After filtering the L1-L2 bonds from the total bonds (Figure 4c), we find that there are 15 bonds being generated after collision. Through the L1-L2 bonds, both layers of graphene have no chance to slide relatively as that happening between two layers of ideal graphene.

Now, we raise two questions: **Can multi-layered graphene ribbon be welded by high-speed fullerenes? And what is the configuration of the new bonds between neighboring or non-neighboring layers after collision?** To answer the questions, we analyze the bond layout at the damaged areas in the ribbon with three or four layers of graphene. Meanwhile, further 300 ps relaxation for the system are fulfilled after 3 ps of collision. The stability of the final configuration of the damaged graphene ribbon is evaluated after full relaxation. **If the three-/four-layered ribbon can be welded via collision with fullerenes, logistically, the thicker ribbon can also be welded by the fullerenes with higher speed.**

3.3. Carbon networks in multi-layered graphene ribbon after collision

*(a) Three-layered ribbon*

When the three-layered graphene ribbon is collided by the two fullerenes with $v$=140 Å/ps (Figure 5a), they have different numbers of broken bonds, e.g., 12 in L1 and 22 in L3, respectively. The numbers of broken bonds are not identical, but both are higher than that in the middle layer L2. Hence, the surface layers are damaged heavier than the middle layer during collision. The snapshots also illustrate that the two fullerenes break with different degrees, which can be verified from the difference of their VPE curves shown in Figure 6a. They are connected with the corresponding surface layers via different number of new bonds. For example, there are 10 to 13 B1-L1 bonds and 14 to 19 B2-L3 bonds. We also find that B1 is not covalently bonded with L2 or L3, and B2 is not bonded with L1 or L2. It means that the two fullerenes do not penetrate the three layers at such incident velocity.

**Although the middle layer L2 has no covalent bond with the two fullerenes, it has new bonds to connect the two surface layers**, e.g., the numbers of L1-L2 and L2-L3 bonds are non-zero. However, the two numbers of bonds are also different due to asymmetric collision. Meanwhile, the two surface layers have no covalent bonds during collision because the fullerenes do not penetrate the whole graphene ribbon.

In Figure 5b, two fullerenes with higher incident velocity of 200 Å/ps penetrate the three-layered ribbon successfully within one pico-second. This can be verified by the covalent bonds generated between B1 and L3 or B2 and L1. For example, at 2900 fs, there are 4 B1-L3 bonds and 2 B2-L1 bonds.

Impacted by such high-speed fullerenes, the three layers of graphene have large number of broken bonds, e.g., at 2900 fs, they subsequently have 61, 97 and 78 broken bonds. Among the three layers, the middle layer has the highest number of broken bonds because it is penetrated by both fullerenes. Hence, two large damaged areas appear in the middle layer after collision, and lead to the highest increasing of potential energy (see L2 curve in Figure 6b). It also means that neighboring layers have chance to form covalent bonds with each other. For example, at 870 fs, there are 35 L1-L2 bonds and 40 L2-L3 bonds.

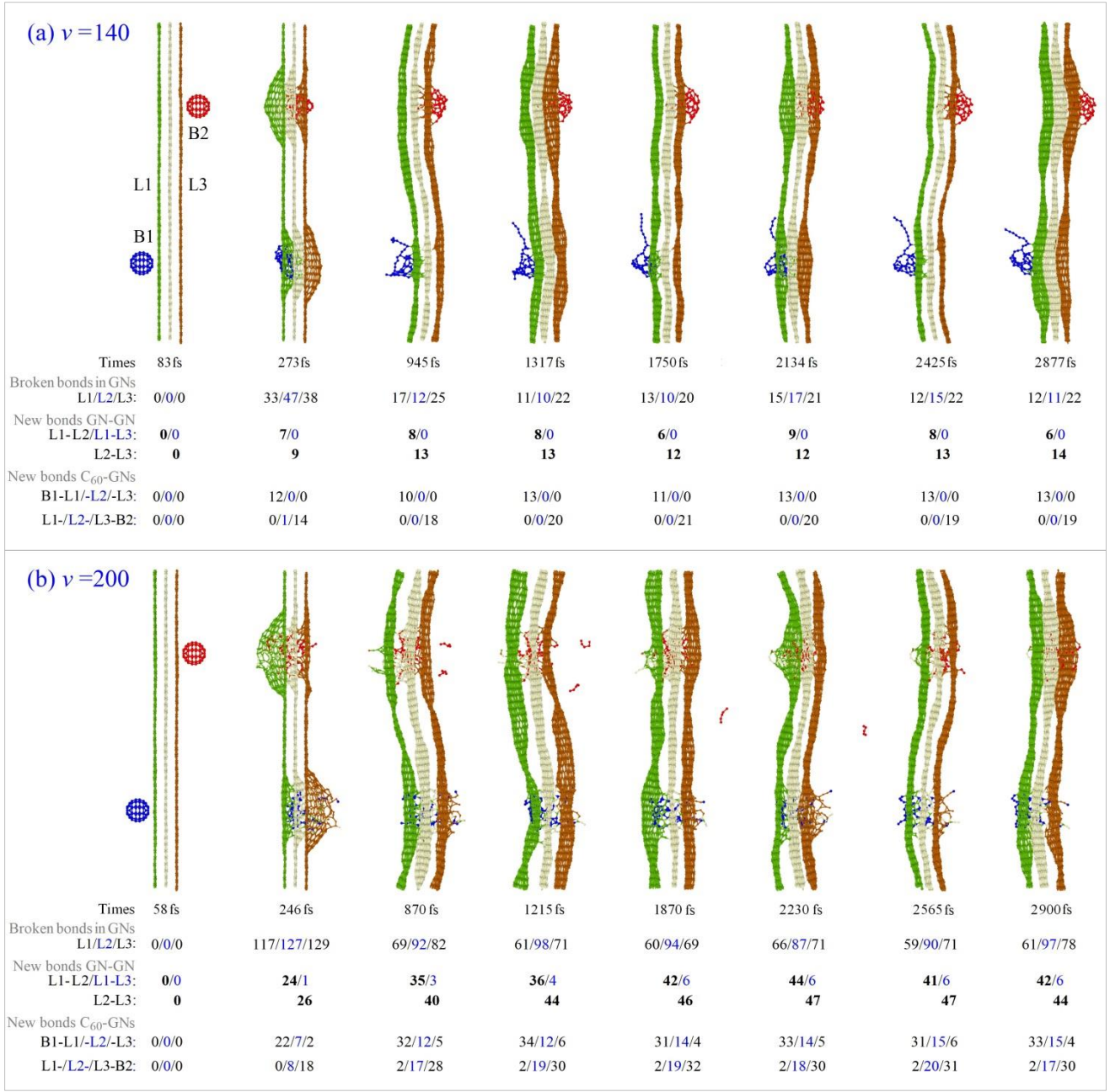

Figure 5  Snapshots of the three-layered graphene ribbons system during collision with incident velocities of (a) $v$=140 Å/ps, (b) $v$=200 Å/ps. The numbers of broken bonds and new bonds between different components at both impact areas are listed. "GN" means graphene.

It is necessary to demonstrate that new bonds between the atoms from both surface layers are also generated after collision. For example, there are 3 L1-L3 bonds at 870 fs, or 6 bonds at 2900 fs. However, the new bonds do not connect the two surface layers with distance of ~7 Å, which is far greater than the cutoff 1.65 Å. During collision, some atoms on L1 are pushed away by B1. They penetrate the middle layer, and are finally captured by L3. Similarly, some atoms on L3 penetrate the middle layer, and are captured by L1. This is the reason for the generation of new L1-L3 bonds during collision.

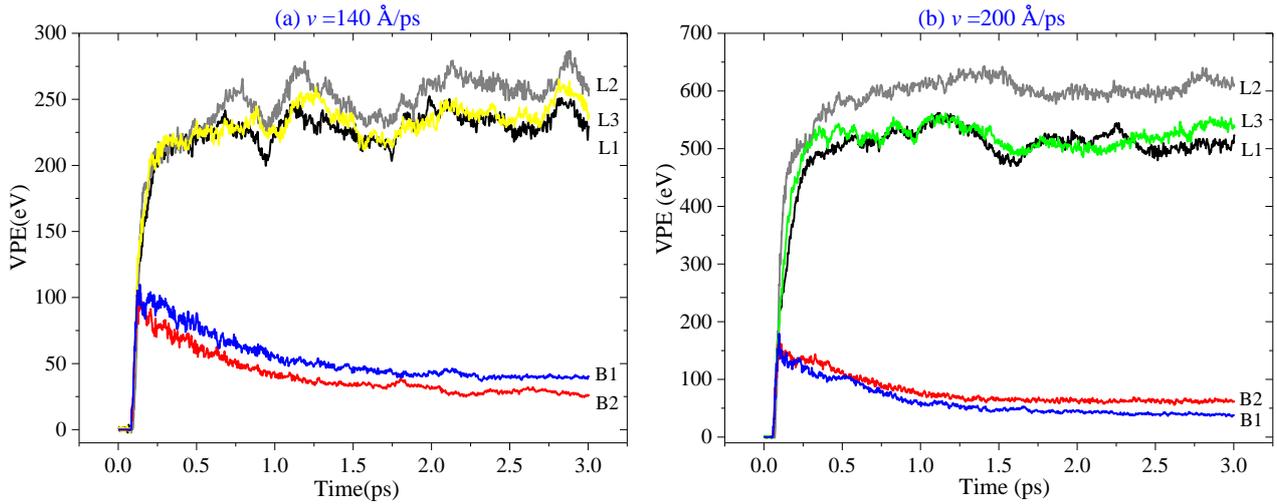

Figure 6  Histories of VPE of the three layers in ribbon and two fullerenes during colliding with (a) *v*=140 Å/ps, (b) *v*=200 Å/ps at 300 K.

*(b) Four-layered ribbon*

When the ribbon contains four layers of graphene, the lower limit of incident velocity for the two fullerenes increases to 160 Å/ps, and the upper limit becomes 260 Å/ps. Impacted by the two fullerenes with speed of 160 Å/ps, the representative snapshots together with the bond variations are given in Figure 7a. Comparing with the three-layered ribbon, the four-layered ribbon is stiffer, and has smaller deformation when they are compacted by the same fullerenes with same speed. Hence, to break the bonds in the layers, the two fullerenes should have higher incident velocity.

During collision, the two surface layers (L1 and L4) have more broken bonds than middle layers (L2 and L3). For instance, at 2679 fs in Figure 7a, the four layers have 33, 26, 17 and 27 broken bonds, respectively. According to the new bonds between $C_{60}$ and graphene ribbons (i.e., $C_{60}$-GNs), both fullerenes penetrate the three layers of graphene. But, B1 have more covalent bonds connecting with graphene than B2. During collision, the two middle layers have higher increasing of potential energy than the two surface layers (Figure 8a). The difference of VPE between the two middle layers or between the two surface layers is not obvious. Due to asymmetric collision, the two fullerenes have different values of VPE at 3 ps. One major reason is that the two damaged fullerenes have different bonding state with the ribbon.

As the graphene ribbon is broken by the two fullerenes, the neighboring layers are bonded together, and carbon network are formed. For example, at 540 fs, there are 17 L1-L2 bonds, 11 L2-L3 bonds and 14 L3-L4 bonds. At 2679 fs (Figure 7a), there is a new bond connecting L2 and L4, which is caused by collision of B2. But the number of L1-L4 bond is always zero. The neighboring layers, e.g., L2-L3, have only 8 new bonds. Hence, the four layers are not connected tightly.

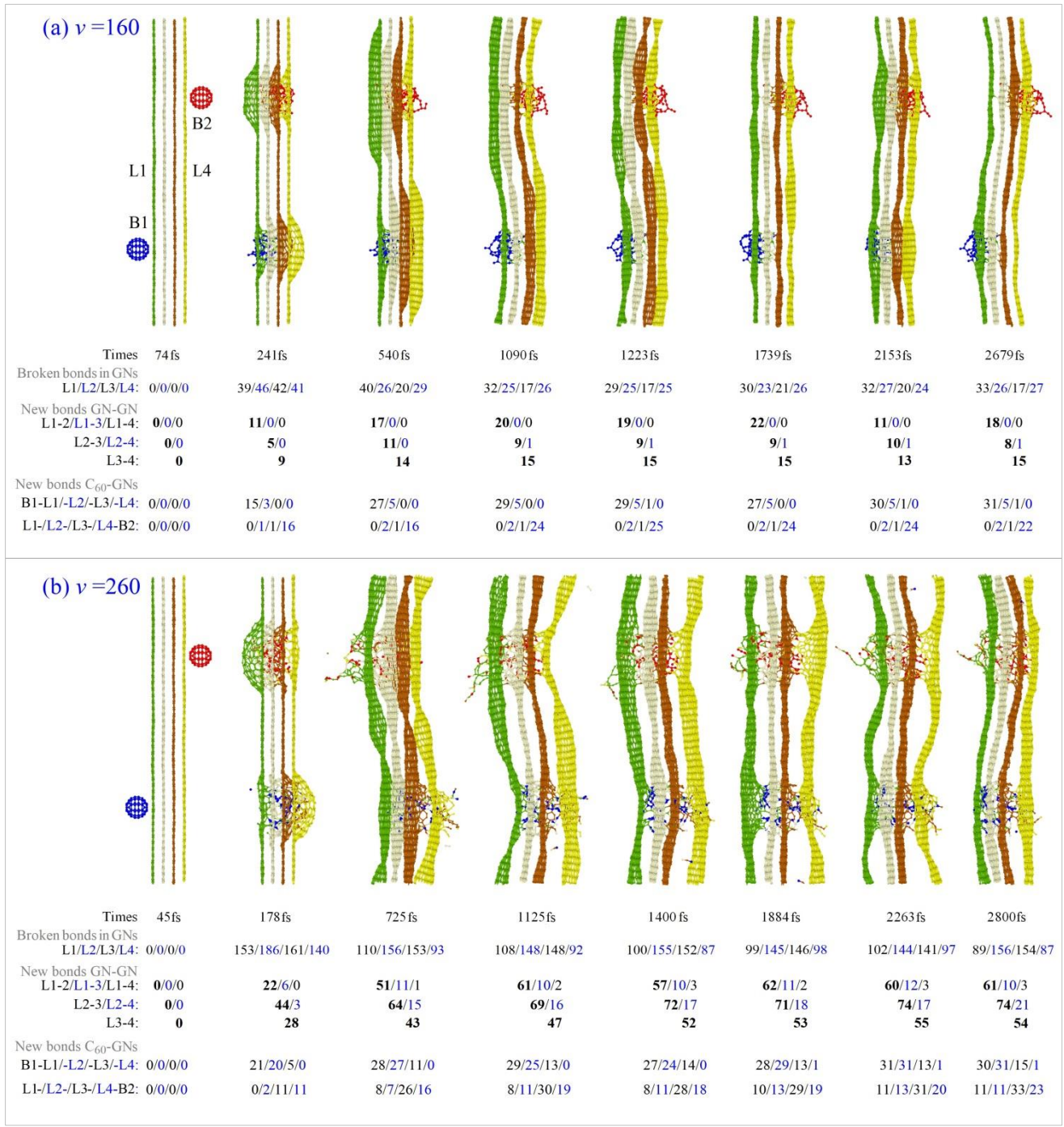

**Figure 7** Snapshots of the four-layered graphene ribbons system during collision with incident velocities of (a) $v$=160 Å/ps, (b) $v$=260 Å/ps. The numbers of broken and new bonds at both impact areas are listed. "GN" means graphene.

**To improve the connection among the four layers of graphene, velocities of the two fullerenes are increased to be 260 Å/ps** (Movie 2, Movie 3). Figure 7b gives some snapshots of the system within 3 ps of collision. At 178 fs, bonds of L1-L3 and L2-L4 have been generated. At 725 fs, bonds L1-L4 appear. At 2800 fs, there are 10 L1-L3 bonds, 21 L2-L4 bonds and 3 L1-L4 bonds. The neighboring layers, e.g., L2 and L3, have more than 50 new bonds, which means that **the four layers are well connected after collision**. In this case, the lower number of L1-L4 bonds does not mean the two fullerenes do not fully penetrate the four layers of graphene. From the snapshot at 725 fs, one can

find that L1 has been penetrated by B2 (red ball), and L4 is penetrated by B1. According to the VPE curves of the four layers in Figure 8b, L2 and L3 have different damage state, e.g., L3 is not well bonded with its neighboring graphene or fullerenes during collision. However, in this case, the VPE curves of B1 and B2 are different slightly. The reason is that they have similar number of new bonds with graphene. For example, at 2800 fs, there are 77(=30+31+15+1) new bonds between B1 and graphene, or 78(=11+11+33+23) new bonds connecting B2 and graphene.

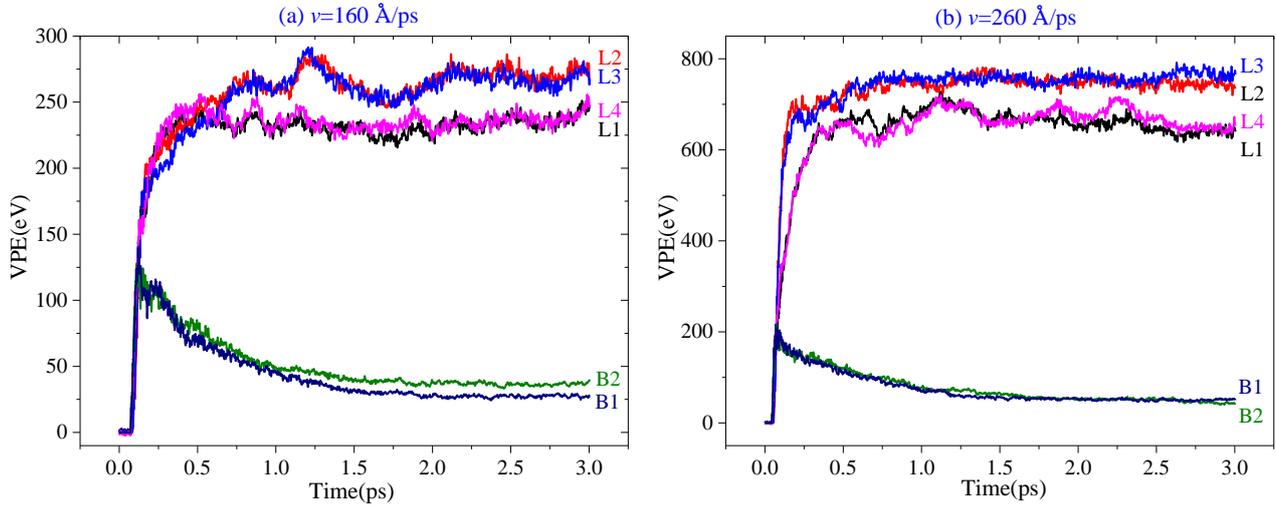

Figure 8   Histories of VPE of the four layers of graphene and two fullerenes during collision with (a) $v$=160 Å/ps, (b) $v$=260 Å/ps.

3.4. Relaxation of graphene ribbons after collision

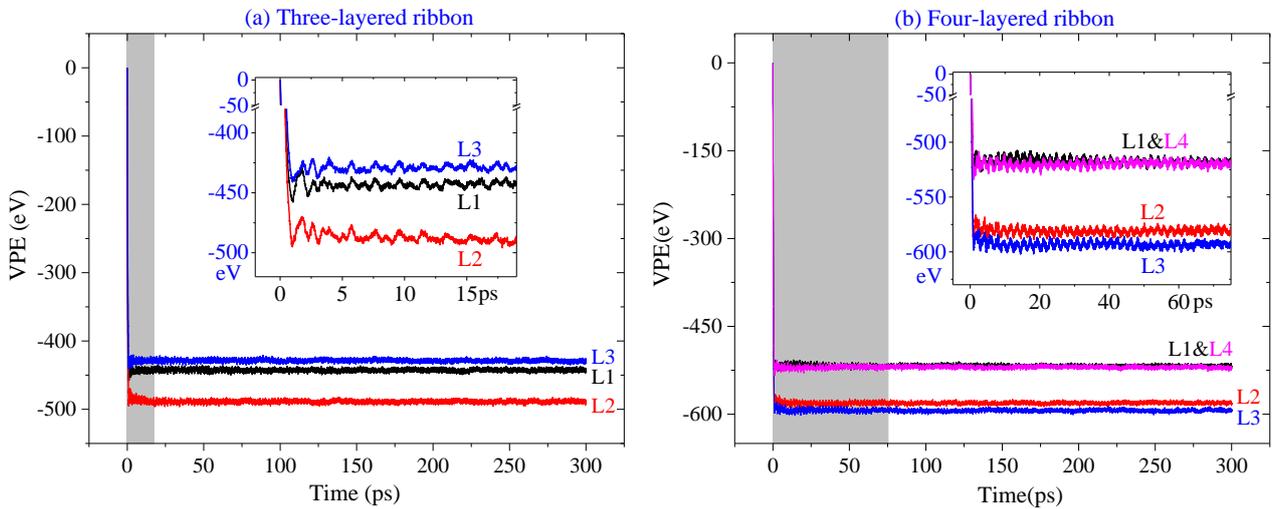

Figure 9   Histories of VPE of different layers in graphene ribbons after collision. (a) The three-layered ribbon collided by the fullerenes with $v$=200 Å/ps. (b) The four-layered ribbon collided by the fullerenes with $v$=260 Å/ps.

After 3 ps of collision, we remove all the atoms that escape from the ribbon and put the remained part in NVT ensemble with temperature of 300 K. After 300 ps of relaxation, the carbon network, i.e., the local layout of new bonds, changes slightly. From the VPE curves shown in Figure 9a, we find that the system tends to be stable within 20 ps for the three-layered ribbon. The small difference between VPE values of the two surface layers, i.e., L1 and L3, is caused by their asymmetric collision with fullerenes. **Even after relaxation, they still have different bond configurations at the two impact areas.** Compared with the two surface layers, the middle layer L2 has larger descend of its potential energy. Similarly, in Figure 9b, the two VPE curves of L1 and L4 almost coincide, and the VPE of L2 and L3 decrease greater than that of L1 and L4.

(a) Three-layered ribbon

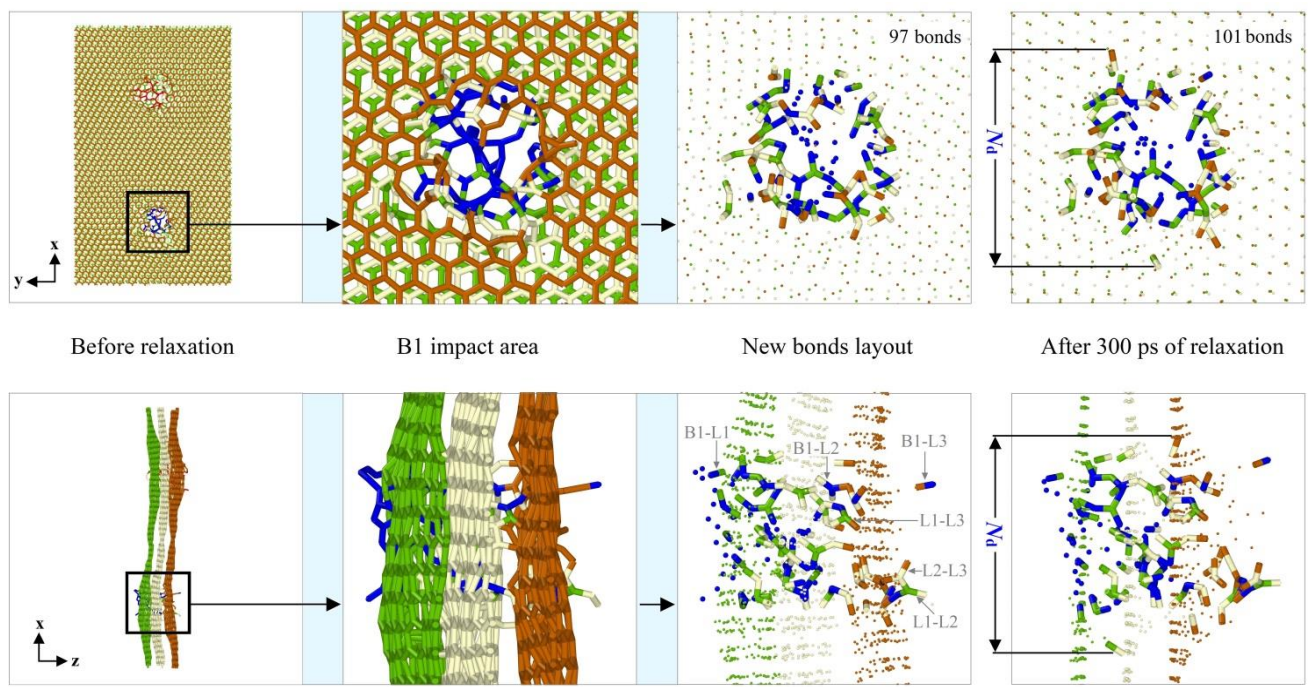

(b) Four-layered ribbon

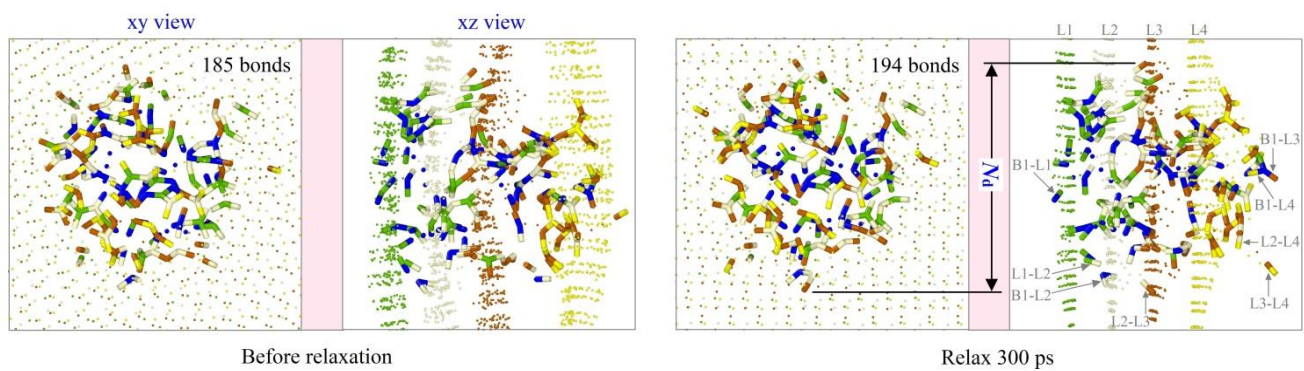

Figure 10 Local new bonds between different layers at the B1 impact area before and after 300 ps of relaxation. $N_d$ is the maximum damage length. (a) On the three-layered ribbon after impacted by the fullerenes with $v$=200 Å/ps. The width of the B1 damage area is $N_d$ =16.69 Å. (b) On the four layered ribbon after impacted with $v$=260 Å/ps. $N_d$ =19.94 Å. The cutoff for bond length is 1.65 Å.

**To show the stability of the network formed between graphene and fullerenes, we draw the bond**

configurations at the B1 impact area on both three-layered (Figure 10a) and four-layered ribbons (Figure 10b). For example, with a cutoff of 1.65 Å, there are 97 new bonds at the B1 impact area in the three-layered ribbon after collision. After relaxation for 300 ps, there are 101 new bonds in the same region, which means that four more bonds being generated during relaxation. As listed in Table 1, two of the four bonds are generated between neighbor layers. When increasing the cutoff to 2 Å, there are only two new B1-L1 bonds being generated after relaxation, and no new bond is generated among the three layers. At 300 ps, the number of new bonds with cutoff of 2.0 Å is five (=106-101) more than that of 0.65 Å, including 4 weaker B1-L1bonds. The number is less than 5% of the total number of new bonds. Hence, the layout of new bonds formed by collision is stable.

In the four-layered ribbon, there are 185 new bonds shorter than 1.65 Å at the B1 impact area after collision (Table 1). Nine bonds are generated after 300ps of relaxation, and seven of them are generated between neighboring damaged layers. After relaxation, there are 7 (=201-194) bonds with lengths between 1.65 Å and 2 Å, i.e., ~3.5% new weaker bonds are generated. Hence, the B1 impact area is mostly covered by strong covalent bonds.

Table 1 Number of new bonds at the B1 impact area in the damaged ribbons before (0 ps) and after (300 ps) relaxation when bond length has different cutoffs.

| | Three-layered ribbon | | | | | Four-layered ribbon | | | |
|---|---|---|---|---|---|---|---|---|---|
| | Cutoff=1.65 Å | | Cutoff=2.0 Å | | | Cutoff=1.65 Å | | Cutoff=2.0 Å | |
| Bond | 0 ps | 300 ps | 0 ps | 300 ps | Bond | 0 ps | 300 ps | 0 ps | 300 ps |
| L1-L2 | 25 | 26 | 26 | 26 | L1-L2 | 40 | 41 | 41 | 41 |
| L1-L3 | 2 | 2 | 2 | 2 | L1-L3 | 6 | 7 | 6 | 7 |
| L2-L3 | 16 | 17 | 18 | 18 | L1-L4 | 1 | 1 | 1 | 1 |
| B1-L1 | 33 | **35** | 37 | **39** | L2-L3 | 26 | 28 | 29 | 29 |
| B1-L2 | 15 | 15 | 15 | 15 | L2-L4 | 7 | 8 | 8 | 8 |
| B1-L3 | 6 | 6 | 6 | 6 | L3-L4 | 28 | 30 | 32 | 31 |
| | | | | | B1-L1 | 32 | 30 | 32 | 33 |
| | | | | | B1-L2 | 30 | 32 | 34 | 34 |
| | | | | | B1-L3 | 14 | 16 | 14 | 16 |
| | | | | | B1-L4 | 1 | 1 | 1 | 1 |
| Total | 97 | 101 | 104 | 106 | Total | 185 | 194 | 198 | 201 |

**4. Conclusions**

To prevent the relative sliding between neighboring layers of a few-layered graphene ribbon and improve its bending stiffness, we adopt high-speed $C_{60}$ fullerenes to collide with the ribbon from both sides at room temperature. Molecular dynamics simulations are used to illustrate the collision effect on the configurations of the ribbons with consideration of the number of graphene layers and the injection velocity of the fullerenes. Results are discussed and some conclusions are drawn as below.

(1) When peening the graphene ribbons with one to four layers, the intervals of injection velocity of $C_{60}$ fullerenes are [115, 135] Å/ps, [130, 165] Å/ps, [140, 200] Å/ps, and [160, 260] Å/ps, respectively. The boundaries of the intervals increase with the number of layers due to support from neighboring layers.

(2) Even impacted by synchronous fullerenes with same speeds, the newly formed carbon networks at the two impact areas are different due to asymmetric collision effect induced by the thermal vibration of atoms on the graphene.

(3) After being impacted by the $C_{60}$ fullerenes with velocity within the interval, the neighboring layers of graphene will be connected via the newly formed carbon network, and they cannot slide relatively anymore.

(4) The newly formed carbon network at impact areas is mostly covered by strong carbon-carbon covalent bonds, which is stable at 300 K.

Using the present peening-like method with high-speed fullerenes, a new 2D carbon material can be fabricated. Besides, the density of welding points (impact areas) and the thickness of the new carbon material can be well controlled, which will benefit its theoretical modeling and applied research, as well as relevant in several areas of technology, such as materials science and engineering, automotive, aerospace, and defense armor.

**Compliance with Ethical Standards:** The authors comply with the ethical rules for this journal.

**Funding:** The authors are grateful for financial support from National Key Research and Development Plan, China (Grant No.: 2017YFC0405102).

**Conflict of Interest:** The authors declare that they have no conflict of interest.

**Data Availability:** There is no datum obtained from the third party.

**Acknowledgements:** Not applicable.

**Supporting Materials:** Movie 1--1L-v=115-during [90, 300]fs.avi; Movie 2--4L-v=260-during [45, 200]fs.avi; Movie 3--4L-v=260-during [45, 200]fs-new bonds at B1 impact area.avi

References

1  Novoselov KS, Geim AK, Morozov SV, et al. Electric field effect in atomically thin carbon films. *Science*, 2004, 306(5696): 666-669

2  Meyer JC, Geim AK, Katsnelson MI, et al. The structure of suspended graphene sheets. *Nature*, 2007, 446(7131): 60-63

3  Zandiatashbar A, Lee GH, An SJ, et al. Effect of defects on the intrinsic strength and stiffness of graphene. *Nature Communications*, 2014, 5: 3186

4  Morozov SV, Novoselov KS, Katsnelson MI, et al. Giant intrinsic carrier mobilities in graphene and its bilayer. *Physical Review Letters*, 2008, 100(1): 016602

5  Cao Y, Fatemi V, Fang S, et al. Unconventional superconductivity in magic-angle graphene superlattices. *Nature*, 2018, 556(7699): 43-50

6  Balandin AA, Ghosh S, Bao WZ, et al. Superior thermal conductivity of single-layer graphene. *Nano Letters*, 2008, 8(3): 902-907

7  Cote LJ, Kim F, Huang JX. Langmuir−blodgett assembly of graphite oxide single layers. *Journal of the American Chemical Society*, 2009, 131(3): 1043-1049

8  Eftekhari A, Jafarkhani P. Curly graphene with specious interlayers displaying superior capacity for hydrogen storage. *The Journal of Physical Chemistry C*, 2013, 117(48): 25845-25851

9  Zhong MY, Xu DK, Yu XG, et al. Interface coupling in graphene/fluorographene heterostructure for high-performance graphene/silicon solar cells. *Nano Energy*, 2016, 28: 12-18

10  Braga SF, Coluci VR, Legoas SB, et al. Structure and dynamics of carbon nanoscrolls. *Nano Letters*, 2004,

          4(5): 881-884

11      Wallace J, Shao L. Defect-induced carbon nanoscroll formation. *Carbon*, 2015, 91: 96-102

12      Aust RB, Drickamer HG. Carbon: A New Crystalline Phase. *Science*, 1963, 140(3568): 817-819

13      Mao WL, Mao HK, Eng PJ, et al. Bonding changes in compressed superhard graphite. *Science*, 2003, 302(5644): 425-427

14      Barboza AP, Guimaraes MH, Massote DV, et al. Room-temperature compression-induced diamondization of few-layer graphene. *Advanced Materials*, 2011, 23(27): 3014-3017

15      Martins LGP, Matos MJ, Paschoal AR, et al. Raman evidence for pressure-induced formation of diamondene. *Nature Communications*, 2017, 8(1): 1-9

16      Wang L, Cai K, Xie YM, et al. Thermal shrinkage and stability of diamondene nanotubes. *Nanotechnology*, 2019, 30(7): 075702

17      Wang L, Cai K, Wei SY, et al. Softening to hardening of stretched diamondene nanotubes. *Physical Chemistry Chemical Physics*, 2018, 20(32): 21136-21143

18      Cai K, Wang L, Xie YM. Buckling behavior of nanotubes from diamondene. *Materials and Design*, 2018, 149: 34-42

19      Chernozatonskii LA, Sorokin PB, Kvashnin AGE, et al. Diamond-like C2H nanolayer, diamane: Simulation of the structure and properties. *JETP Letters*, 2009, 90(2): 134-138

20      Chernozatonskii LA, Sorokin PB, Kuzubov AA, et al. Influence of size effect on the electric and elastic properties of diamond films with nanometer thickness. *Journal of Physical Chemistry C*, 2011, 115(1): 132-136

21      Kvashnin AG, Chernozatonskii LA, Yakobson BI, et al. Phase diagram of quasi-two-dimensional carbon, from graphene to diamond. *Nano Letters*, 2014, 14(2): 676-681

22      Jiang JW, Leng JT, Li JX, et al. Twin graphene: A novel two-dimensional semiconducting carbon allotrope. *Carbon*, 2017, 118: 370-375

23      Yang Y, Cai K, Shi J, et al. Shrinkage-expansion of a tri-isometric knitting from graphene ribbons at finite temperature. *Materials and Design*, 2020, 185: 108269

24      Yang Y, Cai K, Shi J, et al. Nanotextures from orthogonal graphene ribbons: Thermal stability evaluation. *Carbon*, 2019, 144: 81-90

25      Iijima S. Helical microtubules of graphitic carbon. *Nature*, 1991, 354(6348): 56-58

26      Lima MD, Li N, Jung De Andrade M, et al. Electrically, chemically, and photonically powered torsional and tensile actuation of hybrid carbon nanotube yarn muscles. *Science*, 2012, 338(6109): 928-32

27      Kim SH, Haines CS, Li N, et al. Harvesting electrical energy from carbon nanotube yarn twist. *Science*, 2017, 357: 773-778

28      Kroto HW, Heath JR, Obrien SC, et al. C60: Buckminsterfullerene. *Nature*, 1985, 318(6042): 162-163

29      Baughman RH, Galvão DS. Crystalline networks with unusual predicted mechanical and thermal properties. *Nature*, 1993, 365: 735-737

30      Hall LJ, Coluci VR, Galvão DS, et al. Sign change of Poisson's ratio for carbon nanotube sheets. *Science*, 2008, 320: 504-507

31      Xu LQ, Wei N, Zheng YP, et al. Graphene-nanotube 3D networks: intriguing thermal and mechanical properties. *Journal of Materials Chemistry*, 2012, 22(4): 1435-1444

32      Abdol MA, Sadeghzadeh S, Jalaly M, et al. Constructing a three-dimensional graphene structure via bonding layers by ion beam irradiation. *Scientific Reports*, 2019, 9(1): 8127

33      Bai ZT, Zhang L, Li HY, et al. Nanopore creation in graphene by ion beam irradiation: geometry, quality, and efficiency. *ACS Applied Materials and Interfaces*, 2016, 8(37): 24803-24809

| 34 | Sadeghzadeh S, Liu L. Resistance and rupture analysis of single- and few-layer graphene nanosheets impacted by various projectiles. *Superlattices and Microstructures*, 2016, 97: 617-629 |
|---|---|
| 35 | Wang HB, Li N, Xu ZW, et al. Enhanced sheet-sheet welding and interfacial wettability of 3D graphene networks as radiation protection in gamma-irradiated epoxy composites. *Composites Science and Technology*, 2018, 157: 57-66 |
| 36 | Yao WJ, Fan L. The effect of ion irradiation induced defects on mechanical properties of graphene/copper layered nanocomposites. *Metals*, 2019, 9(7): 733 |
| 37 | Geng S, Verkhoturov SV, Eller MJ, et al. Characterization of individual free-standing nano-objects by cluster SIMS in transmission. *Journal of Vacuum Science and Technology B, Nanotechnology and Microelectronics: Materials, Processing, Measurement, and Phenomena*, 2016, 34(3): 03H117 |
| 38 | Geng S, Verkhoturov SV, Eller MJ, et al. The collision of a hypervelocity massive projectile with free-standing graphene: investigation of secondary ion emission and projectile fragmentation. *The Journal of Chemical Physics*, 2017, 146(5): 054305 |
| 39 | Sadeghzadeh S. Computational design of graphene sheets for withstanding the impact of ultrafast projectiles. *Journal of Molecular Graphics and Modelling*, 2016, 70: 196-211 |
| 40 | Sadeghzadeh S. Benchmarking the penetration-resistance efficiency of multilayer graphene sheets due to spacing the graphene layers. *Applied Physics A*, 2016, 122(7): 1-12 |
| 41 | Sadeghzadeh S. On the oblique collision of gaseous molecules with graphene nanosheets. *Molecular Simulation*, 2016, 42(15): 1233-1241 |
| 42 | Sadeghzadeh S. Impact dynamics of graphene nanosheets in collision with metallic nanoparticles. *Scientia Iranica*, 2016, 23(6): 3153-3162 |
| 43 | Eller MJ, Liang CK, Della-Negra S, et al. Hypervelocity nanoparticle impacts on free-standing graphene: a sui generis mode of sputtering. *The Journal of Chemical Physics*, 2015, 142(4): 044308 |
| 44 | Verkhoturov SV, Geng S, Czerwinski B, et al. Single impacts of keV fullerene ions on free standing graphene: emission of ions and electrons from confined volume. *The Journal of Chemical Physics*, 2015, 143(16): 164302 |
| 45 | Verkhoturov SV, Czerwinski B, Verkhoturov DS, et al. Ejection-ionization of molecules from free standing graphene. *Journal of Chemical Physics*, 2017, 146(8): 084308 |
| 46 | Stuart SJ, Tutein AB, Harrison JA. A reactive potential for hydrocarbons with intermolecular interactions. *Journal of Chemical Physics*, 2000, 112(14): 6472-6486 |
| 47 | Golunski M, Postawa Z. Effect of kinetic energy and impact angle on carbon ejection from a free-standing graphene bombarded by kilo-electron-volt C60. *Journal of Vacuum Science and Technology B, Nanotechnology and Microelectronics: Materials, Processing, Measurement, and Phenomena*, 2018, 36(3): 03F112 |
| 48 | Becton M, Zhang LY, Wang XQ. Molecular dynamics dtudy of programmable nanoporous graphene. *Journal of Nanomechanics and Micromechanics*, 2014, 4(3): B4014002 |
| 49 | Liesegang D, Oligschleger C. Spectral modifications of graphene using molecular dynamics simulations. *Journal of Modern Physics*, 2014, 05(04): 149-156 |
| 50 | Hosseini-Hashemi S, Sepahi-Boroujeni A, Sepahi-Boroujeni S. Analytical and molecular dynamics studies on the impact loading of single-layered graphene sheet by fullerene. *Applied Surface Science*, 2018, 437: 366-374 |
| 51 | Signetti S, Taioli S, Pugno NM. 2D material armors showing superior impact strength of few layers. *ACS Applied Materials and Interfaces*, 2017, 9(46): 40820-40830 |
| 52 | Stukowski A. Visualization and analysis of atomistic simulation data with OVITO–the Open Visualization |
| | Tool. *Modelling and Simulation in Materials Science and Engineering*, 2010, 18(1): 015012 |
| 53 | Plimpton S. Fast parallel algorithms for short-range molecular dynamic. *Journal of Computational Physics*, 1995, 117: 1-19 |
| 54 | Jones JE. On the determination of molecular fields. II. From the equation of state of a gas. *Proceedings of the Royal Society of London*, 1924, 106(738): 463-477 |
| 55 | Hoover WG. Canonical dynamics: Equilibrium phase-space distributions. *Physical Review A*, 1985, 31(3): 1695-1697 |
| 56 | Nosé S. A unified formulation of the constant temperature molecular dynamics methods. *Journal of Chemical Physics*, 1984, 81(1): 511-519 |